# Strategic Task Offloading for Delay-Sensitive IoT Applications: A Game-Theory-Based Demand-Supply Mechanism with Participation Incentives

Azadeh Pourkabirian, *Member, IEEE*, Amir Masoud Rahmani, Kai Li, *Senior Member, IEEE,* and Wei Ni, *Fellow Member, IEEE.*

*Abstract*— **Delay-sensitive Internet of Things (IoT) applications have drawn significant attention. Running many of these applications on IoT devices is challenging due to the limited processing resources of these devices and the need for real-time responses. Task offloading can minimize latency by transferring computationally intensive tasks from IoT devices to resource-rich edge servers, ensuring delay and performance guarantees. In this paper, we develop a task-offloading approach for delay-sensitive IoT applications in edge computing environments. Unlike existing schemes, we model the task offloading problem as an economic demand and supply model to achieve market balance. The proposed model avoids under- and over-supply, ensuring the computational resources at edge servers (supply) are allocated in a manner that best meets the processing and computational needs of user devices (demand). Given the multi-agent nature of task offloading involving users and service providers with different preferences and objectives, we design a game-theoretic framework using a Vickrey-Clarke-Groves (VCG) auction. This framework analyzes agent interactions and decision-making processes. Additionally, we develop an incentive mechanism to encourage both parties to participate in the auction. The mechanism maximizes user task offloading to edge servers and motivates edge servers to share their computational resources, achieving profitability for both IoT users and edge servers. Simulations demonstrate our method maximizes social welfare, ensures truthfulness, maintains market balance, and provides latency guarantees for delay-sensitive IoT applications.**
*Index Terms*—**Task offloading, Edge computing, Game theory, Incentive mechanism, Demand and supply, IoT.**

## I. Introduction

The Internet of Things (IoT) consists of a vast network of interconnected devices [1],[2], generating and processing massive amounts of data.

A. Pourkabirian is with the School of Computer Science and Statistics, Trinity College Dublin, Dublin, Ireland. E-mail: a.pourkabirian@tcd.ie.
Amir Masoud Rahmani is with the Future Technology Research Center, National Yunlin University of Science and Technology, Yunlin, Taiwan. E-mail: rahmania@yuntech.edu.tw.
K. Li is with the Real-Time and Embedded Computing Systems Research Centre (CISTER), 4249015 Porto, Portugal. E-mail:kai@isep.ipp.pt.
W. Ni is with Data61, CSIRO, Sydney, NSW 2122, Australia, and the School of Computer Science and Engineering, the University of New South Wales, Kengston, NSW 2073, Australia. E-mail: wei.ni@ieee.org.

This network enables real-time applications [3],[4], such as autonomous vehicles, smart healthcare, and industrial automation, which require low-latency processing and high reliability. Indeed, one of the main requirements of these time-sensitive applications [5]-[7] is short response time, e.g., real-time monitoring systems. However, real-time applications often face high latency, particularly when relying solely on centralized cloud computing for data processing, due to traffic load and network congestion. This latency can hinder the performance of applications requiring instantaneous responses.

Edge computing is a promising technology that addresses this issue by processing data closer to the source through edge servers. IoT devices offload data or computation tasks to high-capacity edge servers located near the source, leveraging their computational power and storage capabilities instead of relying solely on the network core. As a result, the data load sent to the central servers in the network core is minimized, significantly reducing latency. Moreover, distributing computing and data among multiple edge devices increases fault tolerance, as other devices can take over tasks in the event of a failure.

### A. Related Work

Recently, extensive studies have investigated task/data offloading for edge computing [8]-[10]. Some studies [11]-[14] have investigated heuristic schemes to obtain the sub-optimal solutions for task processing offloading in IoT, where tasks are offloaded to the edge server with the highest available computational resources without exploring all possible configurations. The authors of [15], [16] formulated a multi-objective optimization problem to model task offloading in unmanned aerial vehicle (UAV)-enabled edge computing. The dynamic programming technique [17] was employed for task assignment instead of solving a complex mixed-integer programming problem. Nevertheless, this method suffers from high computational complexity and memory usage, making it impractical for large-scale or real-time applications in dynamic and resource-constrained environments. Graph theory [18] was used for task offloading by modeling the system as a graph where nodes represent computing resources and edges denote

.

communication links, enabling efficient task allocation and resource optimization [19] through techniques like shortest path, minimum spanning tree, or graph partitioning algorithms.

In the task offloading process, devices offload tasks to minimize individual costs while competing for edge server resources [20]-[22] This strategic behavior creates a complex interaction among devices [23], where each device aims to optimize its performance, often leading to conflicts over limited resources. Game theory is a powerful tool to model and analyze such interactions, enabling the formulation of resource allocation [24],[25] and task offloading as a competitive or cooperative game [26]. By employing game-theoretic approaches [27],[28], it is possible to design strategies that achieve equilibrium, balance resource usage, and ensure fairness, while also optimizing overall system performance. Some studies [29],[30] have formulated the task offloading problem as an auction competition and developed an auction-based game-theoretic approach to model resource trading among multiple users. They considered multiple resource sellers, such as network operators and service providers, who compete to offer their computing resources to users requiring task offloading.

However, existing task-offloading mechanisms often fail to incorporate a demand and supply model, leading to resource inefficiencies, e.g., under- or over-supply. Additionally, the lack of robust incentive mechanisms results in lower participation rates and reduced competition.

B. Contribution

To address these challenges, this paper proposes a VCG auction game theory-based task-offloading approach. This method ensures efficient resource allocation by maximizing social welfare and incentivizing truthful bidding, making it resilient to cheating. By integrating a demand and supply model, the approach balances user needs with edge server capabilities, achieving market equilibrium and reducing inefficiencies. The inclusion of an incentive mechanism encourages active participation, benefiting both users and service providers. The proposed method is applicable in smart manufacturing scenarios (automotive assembly line), where IoT-enabled devices (e.g., sensors and robotic arms) offload critical tasks such as defect detection to edge servers for low-latency processing, while non-urgent tasks like analyzing long-term equipment wear are managed by the cloud for efficiency.

The contributions of this study are summarized as follows:

- We propose a new task-offloading approach for delay-sensitive applications in IoT edge computing environments. Our approach leverages an economic demand and supply model to accurately model the real-world complexities of market behaviors, addressing the widespread challenges of under- and over-supply in existing studies. Capturing the dynamic interactions between demand fluctuations and supply responses ensures resources are allocated efficiently to meet user requirements while maintaining market balance.
- We formulate multi-agent resource trading between users and edge servers as a VCG-auction game to analyze the interactions between IoT devices and edge servers in task offloading. This game concentrates on maximizing social welfare instead of individual gains. It ensures truthfulness in participants' bidding strategies, i.e., users and edge servers truthfully report their preferences or costs, thereby encouraging honest participation.
- Our model addresses the limitations of current models in the task-offloading market by designing an incentive mechanism. This motivates the maximum number of users to offload their tasks to edge servers and encourages the edge servers to share their computational resources efficiently, ensuring the profitability goals of both IoT users and edge servers. By aligning the incentives of all participants, the model fosters active engagement, and promotes resource utilization.

Extensive simulation results validate the effectiveness of our proposed method. The findings not only affirm the robustness of our approach in terms of latency constraint guarantees but also underscore its superiority in maximizing social welfare, ensuring truthfulness, and maintaining market balance.

The remainder of the paper is organized as follows: We describe the problem statement and system model in Section II. The VCG task offloading auction game model is elucidated in Section III. In Section IV, we design an incentive mechanism for collaborative task offloading. Section V provides simulation results and performance evaluation of the proposed approach. We conclude our work in Section VI.

II. SYSTEM DESCRIPTION

Consider an IoT network comprising a base station as a task dispatcher, $K$ IoT devices i.e., user equipment $\{UE_i\}_{i=1}^{K}$ with $N \geq K$ delay-sensitive computation tasks $\{T_n\}_{n=1}^{N}$ to perform, and $M$ edge servers $\{ES_j\}_{j=1}^{M}$ with the upper bound for the number of task executions, see Fig.1.

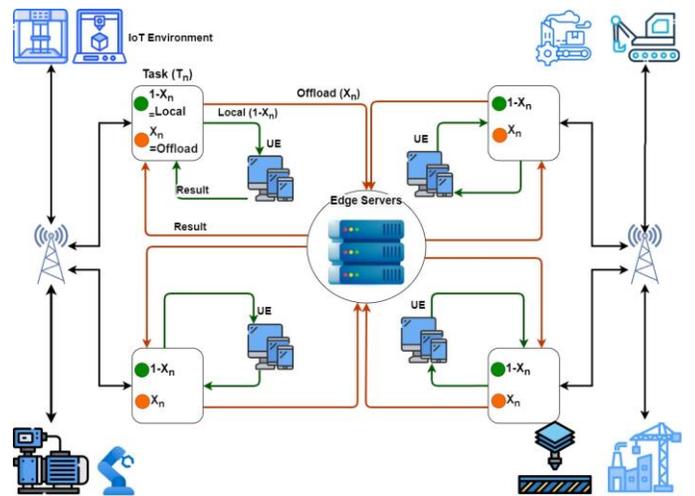

**Fig.1.** A task offloading scenario consisting of base stations, several edge servers, multiple user equipment, and IoT devices performing delay-sensitive applications. User equipment and IoT devices send their task-offloading requests to the BSs. Base stations act as intermediaries, forwarding these requests to edge servers.

These IoT devices exhibit diverse computational requirements such as real-time data processing, complex algorithm execution, and delay constraints. Therefore, they require resource-rich servers to meet the stringent performance and latency demands essential for efficient operation. On the other hand, offloading incurs costs to the UEs, including offloading latency, resource usage fees (resource price), energy consumption for data transmission, and potential security risks. These factors need to be carefully managed to ensure that the benefits of offloading outweigh the associated costs. In this study, UEs take into account trade-offs between delay constraints of their tasks, and resources price, and then send their resource requests to the BS. The BS transmits the tasks' information to nearby ESs. Following the UEs' requests, each ES calculates the amount of computational resources and cost of each resource and then announces them to the BS. The BS coordinates intelligent task scheduling considering the time-sensitivity requirements of tasks and the resource utilization objectives of the ESs. A summary of the primary symbols used in this paper is provided in Table 1.

**Table 1.** The key notation of the problem statement

| Notation | Description |
| --- | --- |
| $K$ | Number of UEs |
| $M$ | Number of ESs |
| $N$ | Number of tasks |
| $T_n$ | Computational task $n$-th |
| $\lambda_{UE}$ | Intensity of UEs |
| $\lambda_{ES}$ | Intensity of ESs |
| $N_{UE}$ | Number of UEs in an ES's coverage area |
| $R$ | The radius of coverage area of an ES |
| $x_n$ | The amount of offloaded task $T_n$ |
| $Len_n$ | The length of the task $T_n$ |
| $O_n$ | The computation complexity of the task $T_n$ |
| $D_n^{max}$ | The delay constraint for the task $T_n$ computation |
| $r_n$ | The transmission rate of the task $T_n$ |
| $W$ | The network bandwidth |
| $N_{sc}$ | The total number of sub-channels |
| $n_{sc}$ | The number of sub-channels allocated to $UE_i$ |
| $P^{tr}$ | The transmission power |
| $\boldsymbol{h}_{kj}$ | The channel coefficients vector between $UE_k$ and $ES_j$ |
| $\sigma^2$ | The variance of the white noise |
| $t^{tr}$ | The transmission time |
| $t_{offload}^{pr}$ | The processing time of an offloaded task $T_n$ |
| $t_{local}^{pr}$ | The local processing time |
| $L_{offload}$ | The offloading latency |
| $L_{local}$ | The local computation latency |
| $L_n$ | The total computation latency of task $T_n$ |
| $C_{total}$ | The total computation cost |

**Assumption 1.** We assume that the UEs and ESs are located according to the heterogeneous Poisson point process (PPP), with intensity $\lambda_{UE}$ and $\lambda_{ES}$, respectively. For simplicity, we consider that the coverage areas of the BS and ESs are circular. Thus, the number of UEs is $N_{UE} = Poisson(\lambda_{UE}\pi R^2)$, $N_{UE} \leq K$ in the circular covered area of radius $R$ (in the coverage area of a typical ES).

The main benefit of selecting the heterogeneous PPP model is that, compared to other models, such as random and uniform distributions where density can be relatively consistent across a given area, the PPP model can easily scale with changes in the density of UEs and ESs. This makes it applicable to various task offloading scenarios, from low-density rural areas to high-density urban areas. The intensities $\lambda_{UE}$ and $\lambda_{ES}$ can be adjusted to reflect different deployment scenarios and densities. In addition, the PPP assumes that the locations of UEs and ESs are independent of each other, which is a reasonable assumption in many practical scenarios where the deployment of UEs and ESs is not coordinated.

In our proposed task offloading scenario, users may choose partial offloading or decide to process tasks completely locally based on resource costs and task offloading latency (i.e., the latency introduced by data transmission and remote processing). If the offloading latency exceeds the task's deadline, the user will decide to process the task locally. Otherwise, the user may offload some parts of the task, ensuring the task is completed before the deadline.

**Assumption 2 (offloading decision).** We model the offloading decision with a Bernoulli distribution, i.e., $b_{i,n} \sim Bernoulli(q_n)$ where $b_{i,n}$ a Bernoulli random variable that $b_{i,n} = 1$ if the UE decides to offload the task $T_n$ and $b_{i,n} = 0$ otherwise. Here, $q_n$ is the probability that the UE decides to offload the task $T_n$.

In general, a task $T_n$ is defined as a profile $\{x_n, Len_n, O_n, D_n^{max}\}$, where $x_n$ represents the amount of the task to be offloaded, $Len_n$ denotes the length of the task $T_n$ (in bits), $O_n$ represents the computation complexity of the task, and $D_n^{max}$ indicates the delay constraint for the task computation. The task is divided into two parts: $(1 - x_n)Len_n$ bits for local computing and $x_n Len_n$ bits for ES computing (i.e., offloading).

**Assumption 3 (partial offloading).** We assume that each computation task $T_n$ can be divided arbitrarily into subsets of varying size, allowing each UE to offload a specific task amount flexibly $x_n \in [0,1]$, to an ES while retaining the remaining task for local processing. If the UE decides to perform the entirety of the computation task $T_n$ locally, $x_n = 0$, whereas if the UE offloads the entire task to an ES, $x_n = 1$.

**Remark 1.** The tasks can still be partitioned flexibly within the constraints imposed by dependencies. Sub-tasks with no dependencies are ideal candidates for offloading or parallel execution. Sequential sub-tasks should ideally be executed on the same resource to reduce communication delays.

In the task offloading scenario, the UEs are constantly seeking ESs for timely task transmission to fulfill their QoS requirement i.e., delay constraint. When a UE offloads a task to an ES, it incurs an offloading cost, including a) the transmission time (i.e., from the UE to the server and vice versa) and b) the computation time of the task at the ES. The transmission time can be calculated as follows:

$$t^{tr}(x_n) \approx 2\frac{x_n Len_n}{r_n}, \qquad (1)$$

where $r_n$ is the transmission rate of the task $T_n$ from $UE_i$ to $ES_j$ and vice versa, which is expressed as follows:

$$r_n = W \log_2\left(1 + \frac{P^{tr}|\mathbf{h}_{ij}|^2}{\sum_{k \neq i} P^{tr}|\mathbf{h}_{kj}|^2 + \sigma^2}\right), \quad (2)$$

where $W$ states the network bandwidth, $P^{tr}$ denotes the transmission power of $ES_j$, $\mathbf{h}_{ij}$ is the channel coefficient vector between $UE_i$ and $ES_j$, $\sum_{k \neq i} P^{tr}|\mathbf{h}_{kj}|^2$ is the aggregated interference from all interfering sources, and $\sigma^2$ is the variance of the additive Gaussian noise. For simplicity, we assume that the transmission rates for both uplink and downlink communications are the same.

The transmission rate of a task plays a significant role in delay constraint satisfaction. Higher data transmission rates allow tasks to be offloaded and results returned more quickly, thereby reducing the overall response time of the offloading process, leading to latency guarantees. To achieve a higher transmission rate, tasks/UEs require more bandwidth. High bandwidth allows for higher transmission rates, reducing the transmission delay significantly.

Given that tasks arrive randomly and independently over time, we model bandwidth allocation using a Poisson point process. This approach helps efficiently distribute resources and optimize network performance.

**Assumption 4.** We assume that the network bandwidth $W$ is divided into $N_{sc}$ sub-channels, allocated to UEs during the offloading process. Based on Assumption 1, the random number of UEs within the coverage area of a serving ES is $N_{UE}$ that require sub-channel allocation for task offloading. The probability of $n_{sc}$ sub-channels being allocated to $UE_i$ at a given time $t$ can be expressed as follows:

$$Pr(N = n_{sc}) = e^{-\lambda_{sc}t} \frac{(\lambda_{sc}t)^{n_{sc}}}{(n_{sc})!}, \quad n_{sc} = 1, \dots, N_s, \quad (3)$$

where $\lambda_{sc}$ is the rate parameter of the Poisson process, which is calculated as follows:

$$\lambda_{sc} = \frac{N_{sc}}{N_{UE}} \mu, \quad (4)$$

in which $\mu$ is the average number of sub-channels requested by a single UE.

As a result, the expected transmission rate for the computational task $T_n$ from $UE_i$ to $ES_j$ can be calculated as

$$\mathbb{E}[r_n] = q_n \left(n_{sc} \left[\log_2\left(1 + \frac{P^{tr}|\mathbf{h}_{kj}|^2}{\sum_{k \neq k, w_k \neq 0} P^{tr}|\mathbf{h}_{kj}|^2 + \sigma^2}\right)\right]\right). \quad (5)$$

On the other hand, the cost associated with the computation time of the task $T_n$ at the $ES_j$, can be calculated as follows:

$$t^{pr}_{offload}(x_n) = \frac{O_n x_n Len_n}{s_j}, \quad (6)$$

where $s_j$ denotes the computational resources of $ES_j$ allocated to the task $T_n$. Consequently, the total offloading latency for the task $T_n$ can be expressed as

$$L_{offload}(x_n) = t^{tr}(x_n) + t^{pr}_{offload}(x_n). \quad (7)$$

It is noteworthy to mention that $UE_i$ processes $(1 - x_n)$ amount of the task, locally on its device. In this case, $UE_i$ incurs a local computing cost in terms of the local computation time $t^{pr}_{local}$ which is expressed as

$$L_{local}(1 - x_n) = t^{pr}_{local}(1 - x_n). \quad (8)$$

Thus, the total computation latency of the task $T_n$, denoted as $L_n$, can be calculated as $L_n = L_{offload}(x_n) + L_{local}(1 - x_n)$, assuming the two segments are not executed simultaneously. If the two segments of the computational task are carried out concurrently, the total computation latency is given by $L_n = max(L_{offload}, L_{local})$.

### A. Demand and Supply Matching Model

Here, we study the task offloading problem in the edge computing environment as a demand and supply-matching model. The proposed model efficiently aligns the UEs' computational demands with available ESs' resources as supplies, ensuring that supply meets user demands without overloading servers. We define a demand and supply profile as $\{b_{i,n}, d_i, a_j, p_j, D_n^{max}\}$ where $d_i$ represents the resource demand of $UE_i$, $a_j$ is the available resource of $ES_j$, and $p_j$ denotes the price associated with a computational resource $s_j$. In our demand and supply model, UEs that generate delay-sensitive tasks, request computational resources from ESs. ESs provide substantial computational resources, referred to as supplies. Each UE makes an offloading decision based on the available resources of an ES, the offloading latency, and the resource price. It decides what proportion of the computing task to offload to the ES and the rest is processed locally to ensure these tasks are processed quickly within specific latency constraints. ESs allocate computational resources to the tasks considering tasks' latency constraints and resource availability. Thus, the total computation cost of a task $T_n$ can be calculated as follows:

$$C_{total}(x_n) = (1 - x_n)(w_1 L_{local}) + x_n(w_1 L_{offload} + w_2 d_{nj} p_j), \quad (9)$$

where $w_1$ is a weighting factor for the latency component, $w_2$ states a weighting factor for the price component, and $d_{nj}$ represents the resource demand of the task $T_n$ from $ES_j$ for the resource $s_j$. The aim is to allocate tasks in a manner that minimizes the offloading cost for each UE while optimizing resource utilization for ESs. In this regard, the optimization problem can be formulated as follows:

$$\min \sum_{j=1}^{M} \sum_{n=1}^{N} x_n(w_1 L_{offload} + w_2 p_j d_{nj}),$$

$$\text{s.t. } C_1: \sum_{j=1}^{M} x_{nj} = 1, \forall j \in ES,$$

$$C_2: \sum_{n=1}^{N} d_{nj} \leq a_j,$$

$$C_3: d_{nj} \geq 0, \forall n \in \{T_n\}, \forall j \in ES,$$

$$C_4: L_n \leq D_n^{max} \quad (10)$$

where constraint $C_1$ guarantees that the entire task is assigned to computational units (an edge server or the local device), and no part of the task is left unassigned or doubly assigned. $C_2$ ensures that the total computational demands of UEs cannot exceed the available resources of an ES. $C_3$ emphasizes the demand of each UE for computational resources must be non-negative. $C_4$ guarantees that the delay constraint of the task $T_n$ is satisfied.

Although the demand and supply matching mechanism considers the requirements of all UEs, ensuring that ESs are neither underutilized nor overloaded, multi-agent frameworks such as game theory can be integrated into this model to take into account the priorities of both parties and incentivize both UEs and ESs to participant in task offloading process. Regarding this, we formulate the task offloading problem as a VCG-auction game to model the competitive behavior in a multi-agent task offloading scenario, considering the preferences and requirements of both UEs and ESs. This model not only enhances fairness in resource distribution but also maximizes social welfare, leading to the satisfaction of both parties.

## III. VCG Task Offloading Auction Game

In the task offloading problem, UEs compete with each other for a serving ES that satisfies their QoS requirements, while ESs tend to serve more UEs to make more income. We formulate these competitions among rational UEs and ESs using a VCG auction game to account for the strategic interactions and preferences of both UEs and ESs. The proposed model promotes fair competition, ensuring that resources are distributed in a manner that benefits the overall system rather than just individual participants.

The proposed VCG task offloading auction game is defined as $G = [K, \{\chi_i\}, \{\pi_i\}]$, where $K$ is the number of UEs as the players of the game, $\chi_i$ denotes the strategy space of $UE_i$, encompassing all possible decisions that UE can make about the amount of resource demands, and $\pi_i$ states the payoff function of $UE_i$.

In our game, the BS acts as an auctioneer and computational resources, e.g., $s_j \in \{S_j\}_{j=1}^M$, are the auctioned items. Each $UE_i$ as a buyer submit a bid vector $bid_i = \{d_i, v_i, D_n^{max}, B_i\}$ to the BS to obtain its required computational resources. The buyers make their bids independently, without knowledge of the other buyers' bids. Here, $v_i$ represents the value that $UE_i$ is willing to pay for the offloading service, i.e., the buyer's suggested price for the computational resource $s_j$ and $B_i$ represents the limited budget of $UE_i$.

The BS collects all UEs' bid vectors and forwards them to the ESs. The ESs, as sellers, then announce the reserve price $p_j$ of each computational resource $s_j$ to the BS in the form of a bid vector $bid_j = \{s_j, p_j, a_j\}$. The BS is responsible for two key issues: selection of winners and resource allocation. It collects bids from participants, calculates payments, selects the winning bidders, determines the optimal allocation of resources based on those bids, and then distributes the resources among the winning bidders. Computation resources are allocated to $N \leq K$ qualified bidders with the highest bids during each time slot $t$. In the event of a tie, where multiple UEs submit identical highest bids, the second criterion i.e., the UE's latency requirement is used to break ties. The resources are socially allocated to the winners, meaning that each bidder obtains only one item. Table 2 provides a summary of the key auction game symbols used in this paper.

**Table 2.** Key notation of task offloading auction game

| Notation | Description |
|---|---|
| $b_i$ | The Bernoulli random variable indicating the offloading decision of $UE_i$ |
| $b_j$ | The Bernoulli random variable indicating the participation of $ES_j$ in the auction |
| $b_{-i}$ | The Bernoulli random variable indicating the participation of other UEs except $UE_i$ in the auction |
| $d_i$ | The resource demand of $UE_i$ |
| $v_i$ | The suggested value of $UE_i$ for the resource $s_j$ |
| $B_i$ | The limited budget of $UE_i$ |
| $s_j$ | The allocated resource of $ES_j$ to $UE_i$ |
| $a_j$ | The available resources of $ES_j$ |
| $p_j$ | The price of the resource $s_j$ |
| $\pi_i^{buyer}(v_i; v_{-i}, \mathbf{s})$ | The payoff function of $UE_i$ |
| $U_i^{buyer}(v_i; v_{-i})$ | The utility function of $UE_i$ |
| $C_i^{buyer}(p_j)$ | The cost function of $UE_i$ |
| $\pi_j^{seller}(\mathbf{s}_j; \mathbf{p}_j)$ | The payoff function of $ES_j$ |
| $U_j^{seller}(\mathbf{s}_j; \mathbf{p}_j)$ | The utility function of $ES_j$ |
| $C_j^{seller}(\mathbf{c}_j)$ | The cost function of $ES_j$ |
| $c_j$ | The cost incurred by the seller to provide the service $s_j$ |
| $pay_i$ | The payment made by each $UE_i$ |
| $l_j$ | The income of $ES_j$ |
| $\beta_i^*(v_{-i}, \mathbf{s})$ | The best response of $UE_i$ |
| $\beta_j^*(s_{ij}, p_j)$ | The best response of $ES_j$ |
| $v_i^*$ | The true bid of $UE_i$ |
| $v_{-i}^*$ | The true bids of others |

**Remark 2.** According to the Assumption 2, the participation of each $UE_i$ in the VCG task offloading auction game can be modeled by Bernoulli distribution, i.e., $b_i \sim Bernoulli(q_i)$, where $q_i$ is the probability that the $UE_i$ participates in the auction and $b_i$ represents a Bernoulli random variable. When $L_{offload} > D_n^{max}$, or $p_j > B_i$, the UE does not participate in the task offloading auction, i.e., $b_i = 0$, otherwise $b_i = 1$.

In the task offloading game, each participant seeks to maximize its payoff. The payoff function of each $UE_i$ can be calculated as follows:

$$\pi_i^{buyer}(v_i; v_{-i}, \mathbf{s}) = \mathbb{E}\left(U_i^{buyer}(v_i; v_{-i})\right) - var\left(C_i^{buyer}(Pay_i(\mathbf{d}))\right), \quad (11)$$

where $\mathbb{E}\left(U_i^{buyer}(v_i; v_{-i})\right)$ is the expected utility that $UE_i$ derives from the offloading service, i.e., the value $\sum_{j \in ES} d_{ij} v_{ij}$ that $UE_i$ expects to gain from the provided $d_{ij}$ demand, and is calculated as follows:

$$\mathbb{E}\left(U_i^{buyer}(v_i; v_{-i})\right) = b_i\left(\sum_{j \in ES} d_{ij} v_{ij}\right). \quad (12)$$

$var\left(C_i^{buyer}(Pay_i(\mathbf{d}))\right)$ represents the variance in the cost incurred by $UE_i$, incorporating both latency cost and user payment. The cost function for $UE_i$ can be expressed as follows:

$$C_i^{buyer}(Pay_i(\mathbf{d})) = b_i\left(w_1 L_{offload} + w_2 Pay_i(\mathbf{d})\right), \quad (13)$$

where $Pay_i(\mathbf{d})$ is the payment that the buyer, $UE_i$, has to be made based on the VCG auction mechanism. According to the VCG auction mechanism, the payment for each winning UE is calculated as the external cost it imposes on other participants. This cost represents the difference in total social welfare with and without the participation of that UE. When a UE participates in the auction and offloads a task to an ES, it uses some of the server's resources, potentially reducing the resources available for other UEs. Essentially, it is the harm or loss in utility that other participants experience because of the inclusion of a particular UE in the resource allocation. This is referred to as external effect, i.e., the impact of one participant's actions on the utility or outcomes of other participants in the auction.

**Definition 1.** Payments made by each UE are inversely related to the external effects they impose on others. This means that when a UE's task offloading greatly reduces the available resources for other UEs, it pays more. Conversely, if the impact of its offloading on the resources available to others is minimal, the payment is lower.

**Lemma 1.** Each buyer (i.e., UE) in the proposed task offloading auction game makes payments that are inversely correlated to the external effects they have on other participants.

*Proof.* Let $S(\boldsymbol{d})$ be the allocation that maximizes the total valuation for the UEs:
$$S(\boldsymbol{d}) \in \arg\max_{\boldsymbol{d}} \boldsymbol{b} \sum_{i=1}^{K}\sum_{j=1}^{M} d_{ij} v_{ij}. \quad (14)$$
According to the VCG mechanism [31], $UE_i$'s payment, $Pay_i(\boldsymbol{d})$, is derived from the difference in total valuations when $i$ is present versus when $i$ is absent. This difference reflects the marginal external effect $UE_i$ imposes on the system, given by
$$\boldsymbol{b} \sum_{\bar{\imath}\in UE,\bar{\imath}\neq i}\sum_{\bar{\jmath}\in S(\boldsymbol{d})} d_{\bar{\imath}\bar{\jmath}} v_{\bar{\imath}\bar{\jmath}} - \boldsymbol{b} \sum_{\bar{\imath}\in UE,\bar{\imath}\neq i}\sum_{\bar{\jmath}\in S(\boldsymbol{d}),\bar{\jmath}\neq j} d_{\bar{\imath}\bar{\jmath}} v_{\bar{\imath}\bar{\jmath}}. \quad (15)$$
When $UE_i$ is present, the total declared valuation is $\boldsymbol{b} \sum_{\bar{\imath}\in UE,\bar{\imath}\neq i}\sum_{\bar{\jmath}\in S(\boldsymbol{d}),\bar{\jmath}\neq j} d_{\bar{\imath}\bar{\jmath}} v_{\bar{\imath}\bar{\jmath}}$, whereas when $UE_i$ is absent, the total declared valuation is $\boldsymbol{b} \sum_{\bar{\imath}\in UE,\bar{\imath}\neq i}\sum_{\bar{\jmath}\in S(\boldsymbol{d})} d_{\bar{\imath}\bar{\jmath}} v_{\bar{\imath}\bar{\jmath}}$.

Therefore, the payment made by each $UE_i$ is inversely related to the external effect it has on the other UEs' valuations. This inverse correlation can be expressed as:
$$Pay_i(\boldsymbol{d}) = f\Big(\sum_{\bar{\imath}\in UE,\bar{\imath}\neq i}\sum_{\bar{\jmath}\in S(\boldsymbol{d})} d_{\bar{\imath}\bar{\jmath}} v_{\bar{\imath}\bar{\jmath}} - \sum_{\bar{\imath}\in UE,\bar{\imath}\neq i}\sum_{\bar{\jmath}\in S(\boldsymbol{d}),\bar{\jmath}\neq j} d_{\bar{\imath}\bar{\jmath}} v_{\bar{\imath}\bar{\jmath}}\Big), \quad (16)$$
where $f(\cdot)$ is a decreasing function indicating the inverse relationship. For simplicity, we ignore $L_{offload}$ in the payment calculation since it has a negligible effect on the payment. So, we can calculate the payment $Pay_i(\boldsymbol{d})$ by each winning $UE_i$ as follows:
$$Pay_i(\boldsymbol{d}) = \boldsymbol{b} \sum_{\bar{\imath}\in UE,\bar{\imath}\neq i}\sum_{\bar{\jmath}\in S(\boldsymbol{d})} d_{\bar{\imath}\bar{\jmath}} v_{\bar{\imath}\bar{\jmath}} - \sum_{\bar{\imath}\in UE,\bar{\imath}\neq i}\sum_{\bar{\jmath}\in S(\boldsymbol{d}),\bar{\jmath}\neq j} d_{\bar{\imath}\bar{\jmath}} v_{\bar{\imath}\bar{\jmath}}. \quad (17)$$
This completes the proof. ∎

Lemma 1 motivates UEs to bid truthfully as the payments align with the true value $\boldsymbol{v}$ they provide to the system. Moreover, it creates a fair and efficient mechanism for resource allocation by making more payments for UEs that impose higher external costs on others. Therefore. Lemma 1 guarantees truthfulness and fairness in our task-offloading method.

In the VCG auction, the aim is to maximize social welfare, ensuring that the overall benefits to participants (i.e., both UEs and ESs) are based on their true valuations. We here define the payoff function for each $ES_j$ as follows:
$$\pi_j^{seller}(\boldsymbol{s}_j;\boldsymbol{p}_j) = \mathbb{E}\Big(U_j^{seller}(\boldsymbol{s}_j;\boldsymbol{p}_j)\Big) - var\Big(C_j^{seller}(\boldsymbol{c}_j)\Big), \quad (18)$$
where $\mathbb{E}\Big(U_j^{seller}(\boldsymbol{s}_j;\boldsymbol{p}_j)\Big)$ is the expected utility function of $ES_j$ and calculated as
$$\mathbb{E}\Big(U_j^{seller}(\boldsymbol{s}_j;\boldsymbol{p}_j)\Big) = b_j \sum_{s_j\in S_j} p_j\, s_j, \quad (19)$$
in which $p_j$ is the reserve price of the resource $s_j$, $s_j = d_{ij}$ based on our demand and supply matching model, $b_j$ represents whether the $ES_j$ participants in the task offloading auction (i.e., $b_j = 1$) or not (i.e., $b_j = 0$). Indeed, when $a_j = 0$ or $c_j > p_j$, $ES_j$ does not participate in the task offloading auction, where $c_j$ represents the cost incurred by the seller to provide the service $s_j$. Here, $var\Big(C_j^{seller}(\boldsymbol{c}_j)\Big)$ represents the variance in the cost incurred by $ES_j$ due to offloading which is defined as
$$C_j^{seller}(\boldsymbol{c}_j) = b_j \sum_{s_j\in S_j} c_j\, s_j. \quad (20)$$

Social welfare is defined as the sum of the payoff of all participants including both the UEs and ESs as follows:
$$Social\ Welfare = \sum_{i\in buyers}\pi_i^{buyer} + \sum_{j\in sellers}\pi_j^{seller}. \quad (21)$$
Our objective is to maximize the overall social welfare. Thus, we formulate the multi-agent task offloading optimization problem as:
$$\begin{aligned}
\max_{v_i, p_j \geq 0} & \sum_{i\in buyers}\pi_i^{buyer} + \sum_{j\in sellers}\pi_j^{seller} \\
\text{Subject to:} \quad & C_1: L_{offload} \leq D_n^{max} \\
& C_2: \sum_{j\in ES} s_{ij} * p_j \leq B_i, \forall i \in UE \\
& C_3: 0 < d_{ij} \leq a_j, \forall i,j \\
& C_4: \sum_{i=1}^{K} d_{ij} \leq S_j, \forall j \in ES \\
& C_5: \sum_{i} d_i \leq \sum_{j} S_j, \forall i,j,
\end{aligned} \quad (22)$$
where constraint $C_1$ emphasizes that the task offloading process should satisfy the delay constraint of each $UE_i$. $C_2$ ensures that the total offloading cost incurred by each $UE_i$ does not exceed its budget. $C_3$ highlights that the demand of each UE must be positive and less than the available resources of the serving $ES_j$. $C_4$ emphasizes each ES does not allocate more resources than it has available. $C_5$ guarantees the overall resource demand from all UEs is less than the overall resource supply from all ESs. This constraint promotes efficient resource management. By keeping the demand within the supply limits, the system remains stable and avoids potential overloads or failures.

In the competitive task offloading game, each UE makes its own best offloading decision in response to its opponents and the ESs' strategies. This decision called the best response, ensures that its payoff is maximized while considering resource constraints and competition from other UEs.

**Definition 2.** For a bid vector $\boldsymbol{bid} = \{bid_1, \ldots, bid_K\}$, a computational resource set $\boldsymbol{s} = \{s_1, \ldots, s_M\}$, and a price set of all shared resources $\boldsymbol{p} = \{p_1, \ldots, p_M\}$, an optimal offloading decision vector for all bidders in the game is given by $\boldsymbol{\beta}^* = \{\beta_1^*, \ldots, \beta_K^*\}$, where $\beta_i^*(v_{-i}, \boldsymbol{s})$ represents the best response of the player $i$ to the decisions of other players and $ES_j$ which is calculated as follows:

$$\beta_i^*(v_{-i}, \mathbf{s}) = \{v_i^* | v_i^* = \arg\max \pi_i^{buyer}(v_i; v_{-i}, \mathbf{s})\}. \quad (23)$$

Similarly, we can extend (23) for best response of each $ES_j$ as

$$\beta_j^*(s_{ij}, p_j) = \{p_j^* | p_j^* = \arg\max \pi_j^{seller}(s_{ij}; p_j)\}. \quad (24)$$

**Theorem 1.** For each $UE_i$, $i \in K$, the best response $\beta_i^*(v_{-i}, \mathbf{s})$ for the task offloading decision exists and has a single value under the following conditions:

$$d_{ij} = \begin{cases} 1, & \text{If } \sum_{j \in ES} v_{ij} \geq \sum_{j \in ES} p_j; \\ 0, & \text{If } \sum_{j \in ES} v_{ij} < \sum_{j \in ES} p_j. \end{cases} \quad (25)$$

*Proof.* We need to show that the payoff function $\pi_i^{buyer}(v_i; v_{-i}, \mathbf{s})$ is concave in $v_i$. A concave utility function ensures that a unique maximum exists. From (11), we need to analyze the concavity of $\mathbb{E}\left(U_i^{buyer}(v_i; v_{-i})\right)$ as follows:

$$\mathbb{E}\left(U_i^{buyer}(v_i; v_{-i})\right) = b_i\left(\sum_{j \in ES} d_{ij} v_{ij}\right). \quad (26)$$

The above function is linear in $d_i, v_i$. A linear function is both concave and convex. Therefore, $\mathbb{E}\left(U_i^{buyer}(v_i; v_{-i})\right)$ is concave in $v_i$. The variance term $var\left(C_i^{buyer}(p_j)\right)$ is typically non-negative and does not affect the concavity of the expected utility function directly. Therefore, the overall payoff function $\pi_i^{buyer}(v_i; v_{-i}, \mathbf{s}) = \mathbb{E}\left(U_i^{buyer}(v_i; v_{-i})\right) - var\left(C_i^{buyer}(p_j)\right)$ remains concave in $v_i$. Since the payoff function is concave in $v_i$, a unique maximum exists. Therefore, the best response $\beta_i^*(v_{-i}, \mathbf{s})$ exists and is unique.

To derive the single value of the best response $\beta_i^*(v_{-i}, \mathbf{s})$ for the task offloading decision in the VCG auction, we need to find $v_i$ that maximizes the payoff $\pi_i^{buyer}(v_i; v_{-i}, \mathbf{s})$. Given (11), we assume the variance term is minimized. Thus, we need to find $d_i$ that maximizes (11). As a result, we can write

$$\pi_i^{buyer}(v_i; v_{-i}, \mathbf{s}) = b_i\left(d_{ij} \sum_{j \in ES} v_{ij} - p_j\right). \quad (27)$$

To find the best response, we take the derivative of the utility function with respect to $v_i$ and set it to zero as follows:

$$\frac{d}{dv_i} \pi_i^{buyer}(v_i; v_{-i}, \mathbf{s}) = \frac{d}{dv_i}\left(b_i\left(d_{ij} \sum_{j \in ES} v_{ij} - p_j\right)\right). \quad (28)$$

Note that $b_i$, $d_{ij}$ and $p_j$ are constants. On the other hand, the term inside the derivative is linear to $v_i$, so we can directly differentiate it. Factoring out $b_i$ and $d_{ij}$, we have

$$b_i d_{ij} \frac{d}{dv_i}\left(\sum_{j \in ES} v_{ij}\right) - \frac{d}{dv_i}(b_i p_j). \quad (29)$$

$p_j$ is also constant, so

$$b_i d_{ij} \frac{d}{dv_i}\left(\sum_{j \in ES} v_{ij}\right) - 0. \quad (30)$$

Given that $\sum_{j \in ES} v_{ij}$ represents a summation over $v_{ij}$ for different $j$, assuming $v_i$ is one of $v_{ij}$, thus, we can write

$$\frac{d}{dv_i}(v_{i1} + \cdots + v_{iM}) = 1. \quad (31)$$

Therefore, the derivative is given by:

$$b_i d_{ij}(1) = b_i d_{ij}. \quad (32)$$

Hence, the derivative of the given expression with respect to $v_i$ is given by:

$$\frac{d}{dv_i}\left(b_i\left(d_{ij} \sum_{j \in ES} v_{ij} - p_j\right)\right) = b_i d_{ij}. \quad (33)$$

We need to find the conditions under which $d_{ij}$ maximizes the payoff. Since $d_{ij}$ is a linear proportion and assuming the payoff increases linearly with $d_{ij}$ Until the constraints are reached, the best response will be the extreme value. We then define the boundary conditions so that each $UE_i$ offload its entire task, i.e., $d_{ij} = 1$, if the value it places on the resources is greater than or equal to the combined prices of the resources ($\sum_{j \in ES} v_{ij} \geq \sum_{j \in ES} p_j$), and does not offload any part of its task $d_{ij} = 0$, otherwise ($\sum_{j \in ES} v_{ij} < \sum_{j \in ES} p_j$).

Consequently, the single value of the best response in (25) is obtained. This completes the proof. ∎

**Remark 3.** The best response function satisfies the constraints $L_n \leq D_n^{max}$ and $d_i \in [0,1]$ since the best response $\beta_i^*$ is achieved by finding the value of $d_i$ that maximizes the difference between the benefit $U_i^{buyer}(v_i; v_{-i})$ and the cost $C_i^{buyer}(p_j)$. These constraints ensure that the chosen offloading amount is within feasible bounds.

One of the challenges is small participation in the auction. With fewer participants, there is less competition, leading to inefficient resource allocation and under-utilization of ESs.

A sharing-incentive mechanism can increase the number of participants in the auction by motivating both UEs and ESs to contribute to the task-offloading process. This mechanism encourages ESs to share all their available computational resources and enhances the inclination of UEs to offload their tasks. This collaborative approach ensures that the computational demands of UEs are efficiently met, optimizing resource utilization and maximizing overall system performance. As a result, there is greater competition and higher overall social welfare, ultimately resulting in a more efficient and effective task-offloading system.

## IV. INCENTIVE MECHANISM FOR COLLABORATIVE TASK OFFLOADING

We designed an incentive-sharing mechanism to encourage both parties to participate in the auction. Within our model, the UEs initiate the offloading process while offering necessary incentives to the ESs. They are required to pay for their task offloading to the ESs, but they can reduce their payment obligations based on the size of the offloaded task. The ESs, in turn, offer a discount i.e., reduced price to the UEs based on the volume of computational tasks offloaded. This means that the UEs can enjoy a cost advantage in utilizing the ESs' computational resources if they offload larger amounts of tasks. The discount is proportional to the size of the task being offloaded, so the UEs benefit more economically when dealing with larger task sizes. This strategy incentivizes the UEs to offload tasks and use the ESs' resources. Simultaneously, the ESs seize the opportunity to promote their income by serving more tasks and efficiently allocating their computational resources in task processing. This dual benefit structure not only motivates the UEs to optimize their payment outlays by utilizing the ES resources but also ensures the ES's strategic use of available resources for enhanced profitability and resource allocation efficiency.

**Remark 4.** Using the VCG auction game theory principles that ensure truthful reporting of resource needs, avoids collusion [32] in the incentive mechanism of collaborative task offloading.

Suppose that $S_j = \{s_{j1}, ..., s_{jN}\}, N \leq K$ be the allocation vector of $ES_j$ to winner UEs in the offloading game. We define an incentive mechanism for buyer $i$ as follows:
$$Pay_i(d) = b \sum_{\bar{\iota} \in UE, \bar{\iota} \neq i} \sum_{j \in S(d)} (1 - \lambda d_{ij}) v_{\bar{\iota}j} - \sum_{\bar{\iota} \in UE, \bar{\iota} \neq i} \sum_{\bar{J} \in S(d), \bar{J} \neq j} (1 - \lambda d_{\bar{\iota}\bar{J}}) v_{\bar{\iota}\bar{J}}, \quad (34)$$
where $\lambda$ is an incentive factor (i.e., discount factor) for larger tasks. Consequently, the income of $ES_j$, i.e., $I_j$, from allocating the resource vector $S_j$ to all winner UEs, is calculated as
$$I_j = \sum_{i \in UE} Pay_{ij}(d). \quad (35)$$

**Definition 3.** The allocation vector $s$ follows the sharing-incentive mechanism if $\pi_j^{seller}(s, p) \geq 0$ and $\pi_j^{seller}(s, p) \geq \pi_j^{seller}(\bar{s}, p), \forall j$, where $\bar{s}$ is an allocation vector based on another allocation strategy (e.g., Lotteries, markets, barter, rationing, or redistribution of income).

It is worth noting that the sharing-incentive mechanism can be similarly defined for UEs in the form of $\pi_i^{buyer}(v, s)$.

**Lemma 2.** The proposed task offloading approach provides the sharing-incentive property.

*Proof.* We prove the property for sellers first and then extend it to buyers. We need to demonstrate that the payoff derived from the proposed mechanism, $\pi_j^{seller}(s, p)$, is non-negative and at least as large as the payoff derived from any alternative allocation strategy, $\pi_j^{seller}(\bar{s}, p)$. From (18), $\pi_j^{seller}(s, p) = \mathbb{E}\left(U_j^{seller}(s, p)\right) - var\left(C_j^{seller}(s, c)\right)$ is the payoff of $ES_j$ under the proposed task offloading mechanism in which $U_j^{seller}(s, p) = b_j \sum_{s_j \in S_j} p_j s_j$ and $C_j^{seller}(s, c) = b_j \sum_{s_j \in S_j} c_j s_j$. By factoring $s_j$, we have
$$\pi_j^{seller}(s, p) = \sum_{s_j \in S_j} s_j (p_j - c_j) = s \sum_{s_j \in S_j} (p_j - c_j). \quad (36)$$
Without loss of generality, we consider $p_j = I_j$. Suppose that $\bar{s}$ is an alternative allocation strategy, so, the payoff under the incentive-sharing mechanism for this strategy is given by:
$$\pi_j^{seller}(\bar{s}, p) = \bar{I}_j - \bar{c}_j, \quad (37)$$
where $\bar{I}_j = \overline{Pay_i}$ and $\bar{c}_j$ identify the income and cost, respectively, under the alternative strategy $\bar{s}$. We need to prove

$$\pi_j^{seller}(s, p) \geq 0 \text{ and } \pi_j^{seller}(s, p) \geq \pi_j^{seller}(\bar{s}, p), \forall j. \quad (38)$$
According to the general rule in the VCG auction, the income of the ESs, $\bar{I}_j$, must be greater than or equal to the operational cost to an ES participant in the auction. Using (35), we can write:
$$\sum_{i \in UE} Pay_i \geq c_j. \quad (39)$$
As a result, we have
$$\pi_j^{seller}(s, p) = \sum_{i \in UE} Pay_i - c_j \geq 0. \quad (40)$$
It was found that the payoff derived from the proposed mechanism is non-negative. Now, we need to show that the payoff is at least as large as the payoff from any other strategy. Using (38) and (40), we can write
$$\pi_j^{seller}(s, p) = \sum_{i \in UE} Pay_i - c_j \geq \pi_j^{seller}(\bar{s}, p) = \bar{I}_j - \bar{c}_j. \quad (41)$$
Since the operational cost $c_j$ remains constant irrespective of the strategy, i.e., $\bar{c}_j = c_j$, we have
$$\pi_j^{seller}(s, p) = \sum_{i \in UE} Pay_i - c_j \geq \bar{I}_j - c_j = \pi_j^{seller}(\bar{s}, p). \quad (42)$$
On the other hand, we have $\sum_{i \in UE} Pay_i \geq \bar{I}_j$, therefore
$$\pi_j^{seller}(s, p) \geq \pi_j^{seller}(\bar{s}, p), \forall j. \quad (43)$$
This completes the proof. ∎

In IoT environments with diverse and competing demands, it is important to ensure that all UEs receive a fair share of resources based on the specific demands of each task. In this case, all participants feel equitably treated and are satisfied with their resource allocation relative to others. Therefore, envy-freeness is achieved. The proposed task offloading auction approach guarantees utility fairness for the UEs. It creates an envy-freeness allocation where each UE favors its allocation over that of any other UEs, meaning that no UE feels jealous since they all believe their portions are at least as good as those of the other UEs.

**Definition 4.** The allocation $s$ is envy-free, given the bid vector $v$ of UEs if
$$\pi_i^{buyer}(s|v) \geq \pi_i^{buyer}(\bar{s}|v), \forall i \in UE, \quad (44)$$
where $\pi_i^{buyer}(s|v)$ denotes the payoff of $UE_i$ for the allocation $s$ given the bid vector $v$, whereas $\pi_i^{buyer}(\bar{s}|v)$ states the payoff of $UE_i$ for the allocation $\bar{s}$ (the allocation given to another) under the same bid vector. This means that no UE prefers the allocation received by another UE over its own, assuming the bid vector remains unchanged.

The solution of the task offloading game is the auction equilibrium in which demand and supply are balanced (market balance). At the auction equilibrium point, all bidders, i.e., the UEs and ESs, have selected their best response $\beta_i^*$ to the decision of the other bidders, and no bidder is motivated to alter its own decision given the decision of the other bidders because it cannot improve its payoff, see Algorithm 1.

**Definition 5.** The auction equilibrium is defined by truthful bidding $v^*$ where no bidder (e.g., UE or ES) has an incentive to deviate from its chosen strategy unilaterally $v_i^*$ to a different strategy $v_i$, given that the strategies of the other bidders remain fixed $(v_{-i}^*)$ as follows:
$$\pi_i^{buyer}(v_i^*, v_{-i}^*, s) \geq \pi_i^{buyer}(v_i, v_{-i}^*, s), \forall v_i \in \chi_i, v_i \neq v_i^*, \quad (45)$$
in which $v_i^*$ is the truthful bid of the bidder $i$ and $v_{-i}^*$ is the truthful bid of the others.

**Algorithm 1.** Task offloading auction game
1. **Input** $K, M, N, R, W, N_{sc}$.
2. **Output** Auction equilibrium $v^*$.
3. **For each computational task** $T_n$ of $UE_i$
4.    Initialize $\{x_n, Len_n, O_n, D_n^{max}\}$.

```
5.    End for
6.    For each UE_i
7.        Calculate L_offload using Eq.7
8.    End for
9.    For each ES_j
10.       Initialize {s_j, a_j, p_j}.
11.   End for
12.   While (auction equilibrium is achieved)
13.       For each UE_i
14.           bid_i = {d_i, v_i, D_n^max, B_i}
15.           If L_offload > D_n^max, or p_j > B_i then
16.               b_i = 0.
17.           Else
18.               b_i = 1.
19.           Choose v_i* = β_i*(v_-i, s)
20.           Calculate π_i^buyer(v_i*, v_-i*, s) using Eq.11
21.           If π_i^buyer(v_i*, v_-i*, s) ≥ π_i^buyer(v_i, v_-i*, s)
22.               Update v* = [v_1*,...,v_K*]
23.       End for
24.   End while
```

The computational complexity of the algorithm is primarily driven by the bid collection, $O(K)$, bids evaluation to determine the optimal allocation that maximizes the social welfare, $O(K \log K)$, winner determination, $O(K)$ and payment calculation, $O(K^2 \log K)$. Thus, the total computational complexity of the proposed method, considering the additional demand and supply model and incentive mechanism, is $O(K^2 \log K)$.

**Theorem 2.** The proposed task offloading auction game has a unique auction equilibrium in which each bidder truthfully reveals its valuation for the offloading service, and this equilibrium is Pareto-optimal and envy-free.

*Proof.* Let $S(\boldsymbol{d})$ be an allocation that maximizes the total declared valuation:
$$S(\boldsymbol{d}) \in \arg\max_{\boldsymbol{d}} \boldsymbol{b} \sum_{i=1}^{K} \left( \sum_{j=1}^{M} d_{ij} v_{ij} - Pay_i(\boldsymbol{d}) \right). \quad (46)$$
From (17), the payment $Pay_i(\boldsymbol{d})$ by each winning $UE_i$ can be written as follows:
$$Pay_i(\boldsymbol{d}) = \boldsymbol{b} \sum_{\bar{\imath} \in UE, \bar{\imath} \neq i} \sum_{\bar{\jmath} \in S(\boldsymbol{d})} d_{\bar{\imath}\bar{\jmath}} v_{\bar{\imath}\bar{\jmath}} - \sum_{\bar{\imath} \in UE, \bar{\imath} \neq i} \sum_{\bar{\jmath} \in S(\boldsymbol{d}), \bar{\jmath} \neq j} d_{\bar{\imath}\bar{\jmath}} v_{\bar{\imath}\bar{\jmath}}. \quad (47)$$
Since each bidder makes its best response decision at the equilibrium, $s_j = d_{ij}$. Thus, the payoff of each $UE_i$ can be expressed as below:
$$\pi_i^{buyer}(v_i; v_{-i}, \boldsymbol{d}) = b_i \left( \sum_{j \in ES} d_{ij} v_{ij} - Pay_i(\boldsymbol{d}) \right). \quad (48)$$

Given the VCG mechanism [31], the existence of equilibrium follows directly from the properties of the mechanism, which ensures that truthful bidding $v_i^*$ is a dominant strategy as

$$v_i = v_i^*, \forall i \in UE. \quad (49)$$
As a result, a unique and stable allocation and payment structure (i.e., equilibrium point) is achieved where the payoff for each UE is maximized as follows:
$$\pi_i^{buyer}(v_i^*, v_{-i}^*, \boldsymbol{d}) \geq \pi_i^{buyer}(v_i, v_{-i}^*, \boldsymbol{d}), \forall v_i \in \chi_i, v_i \neq v_i^*. \quad (50)$$
The allocation $S(\boldsymbol{d}^*)$ maximizes the sum of the valuations, ensuring that no other allocation can make any UE better off without making another UE worse off. Thus, the equilibrium is Pareto-optimal.

On the other hand, since each UE bids truthfully and pays a price based on the marginal impact of its presence on other UEs, no UE prefers the allocation received by another UE over its own. For any two UEs $i$ and $j$, we write
$$\pi_i^{buyer}(v_i^*, v_{-i}^*, \boldsymbol{d}) \geq \pi_i^{buyer}(v_j, v_{-i}^*, \boldsymbol{d}). \quad (51)$$
This ensures that the equilibrium is envy-free. This completes the proof. ∎

**Remark 5.** The two following conditions should be satisfied to achieve the auction equilibrium:

*C₁ (user fulfillment condition):* for a given optimal resource price vector $\boldsymbol{p}^* = \{p_1^*, ..., p_M^*\}$, every $UE_i$ obtains the required optimal resource $s_j^*$ so that we have
$$\boldsymbol{s}^* = \{s_1^*, ..., s_M^*\} \in \arg\max \pi_i^{buyer}(v_i^*, v_{-i}^*, \boldsymbol{s}^*) \quad (52)$$
*C₂ (server fulfillment condition):* All available resources have been allocated.

The condition $C_1$ makes sure that the equilibrium resource allocation maximizes UEs' payoff for the equilibrium pricing $\boldsymbol{p}^*$ under the UE budget constraint. $C_2$ guarantees the profit maximization of ESs since their resources have been completely sold, and prices are non-negative.

## V. SIMULATION RESULTS

### A. Parameter Setting

In this section, we investigate the numerical results to evaluate the performance of the proposed approach. We conduct simulations under a widely-used dataset called Grid Workload Archive (GWA) [33] to validate the effectiveness of proposed solutions. We assess the performance of our approach under two logs available in the GWA named "DAS-2" and "SHARCNET", and then compare it with two state-of-the-art methods named RACO [26] and RALA [27]. The network scenario involves four edge servers, each of which has limited available resources, randomly distributed across a region measuring 1900 m × 1900 m, one BS centrally located within this area, and 115 UEs that run [1000,10000] different tasks. We assume that the tasks follow a Poisson process arrival with an average arrival interval of 2 tasks per second. The processing (computation) time for UEs, $t_{local}^{pr}$, is uniformly and randomly selected from the range of [3, 8] seconds, and for each ESs, $t_{offload}^{pr}$, is selected from the range of [1, 4] seconds. The normalized resource availability of each ES is drawn from a uniform and random distribution within the range of [0, 1]. The unit cost $p_j$ of computation resources $s_j$ of the ESs is randomly generated from [0.1, 1.0] using a uniform distribution. Additionally, we generate the user bids between 30 and 500 for different ESs. Given that the number of jobs in the log files is large, we use parts of these jobs as the tasks in our simulations. The number of tasks is varied from 1000 to 10,000. Table 3 summarizes the key parameter settings in this scenario.

**Table 3.** Key simulation parameters setting

| Notation | Values |
|---|---|
| K | 115 |
| M | 6 |
| N | [1000,10000] |
| λ | 2 tasks/second |
| $N_{UE}$ | 18 |
| $R$ | 500 m |
| $L_i$ | [1,8] Mb |
| $s_j$ | 32 cores |
| $d_i$ | [1,4] cores |
| $\alpha_j$ | [0,1] |
| $p_j$ | [0.1,1.0] |
| $D_n^{max}$ | [10,75] ms |
| $W$ | 10 MHz |
| $N_{sc}$ | 100 |
| $n_{sc}$ | [1,4] |
| $r_n$ | [1,10] Mbps |
| $P^{tr}$ | 35 dB |
| $\sigma^2$ | -174 dBm/Hz |
| $t_{local}^{pr}$ | [3,8] |
| $t_{offload}^{pr}$ | [1,4] |

*B. Social Utility*

Fig. 2 investigates the social welfare in all benchmarks for the default settings of log files. We first analyze the performance of methods under the DAS-2 log file. The results indicate that the proposed approach achieves the highest social welfare, outperforming the RACO algorithm by approximately 26% and the RALA algorithm by about 38 percent. We analyze the performance of the three algorithms when we change the setting based on the SHARCNET log file in Fig. 3. Our method generates at least double the social welfare compared to the two other algorithms.

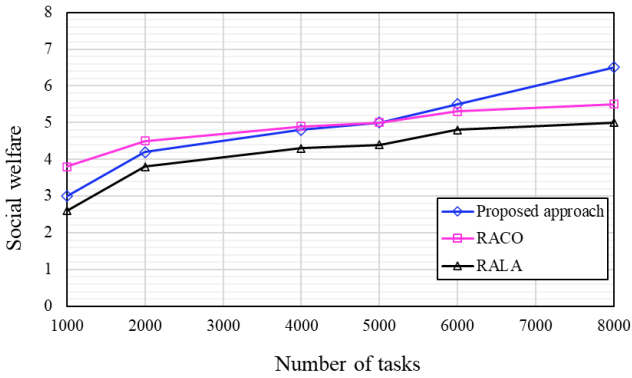

Fig. 2. Social utility for different numbers of tasks under the DAS-2 setting

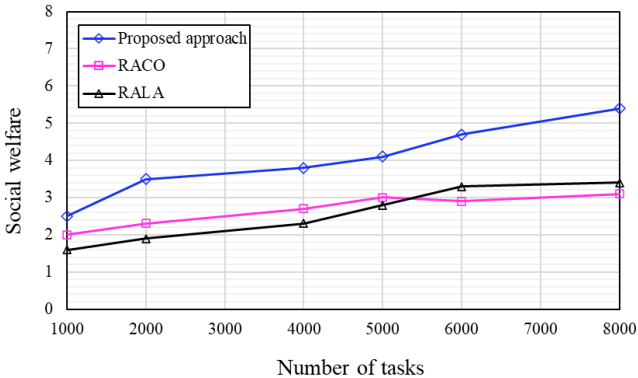

Fig. 3. Social utility for different numbers of tasks under the SHARCNET log file

In Fig. 4, we evaluate how varying the number of UEs impacts the overall social welfare achieved by all methods. As the number of UEs increases, the social welfare in our approach exhibits notable growth. The results demonstrate that as more tasks are introduced, the auction mechanism efficiently allocates available computation resources to the tasks with the highest bids, thus enhancing total social welfare. In the RACO scheme, the ESs compete to offer the lowest price for processing tasks. This price-centric approach often prioritizes cost reduction over optimal resource allocation, which can lead to suboptimal task placement. Consequently, tasks might not be assigned to the servers that value them most highly, resulting in lower overall social welfare. Furthermore, the iterative nature of the RALA method leads to strategic behavior by participants, such as holding back on bids or attempting to game the system, which can distort the auction outcomes and further reduce social welfare. This evaluation validates the robustness and scalability of our method for practical applications in edge computing environments.

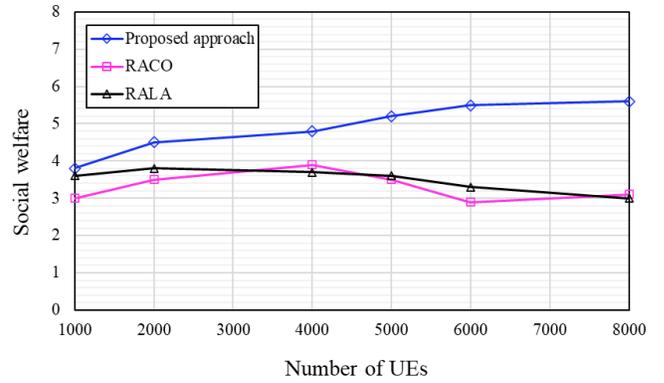

Fig. 4. Social utility under different numbers of UEs

*C. Truthfulness performance*

In Fig. 5, we validate the truthfulness of our algorithm in the SHARCNET setting. In this validation, we allow a UE to submit a bid that is different from its true valuation. The true valuation is fixed at 150. We then vary the submitted bids from 30 to 300. The results demonstrate that both utility values are relatively low, 0.4, when the submitted bid is lower than the true valuation, and they remain unchanged when the submitted bid is larger than the true valuation. This confirms the truthfulness property of our algorithm under the different bid values.

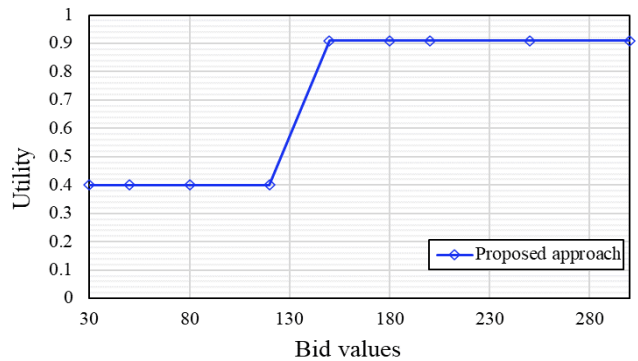

Fig. 5. Truthfulness performance of the proposed method for different bid values under the SHARCNET log file

### D. Rationality analysis

Fig. 6 investigates the individual rationality of UEs by comparing the received payoffs under true bids versus misreported bids using the default settings of the DAS-2 log file. The results demonstrate that users who submitted true bids consistently achieved higher utility compared to those who submitted exaggerated (overbid) or deflated (underbid) bids. The true bids align the users' valuations with the actual costs and optimal utility outcomes. This alignment incentivizes truthful bidding as users realize that any deviation from their true valuation would result in a lower payoff. Consequently, the utility derived from true bids validates the theoretical advantage of the proposed method in promoting honesty and maximizing overall payoff within the auction framework.

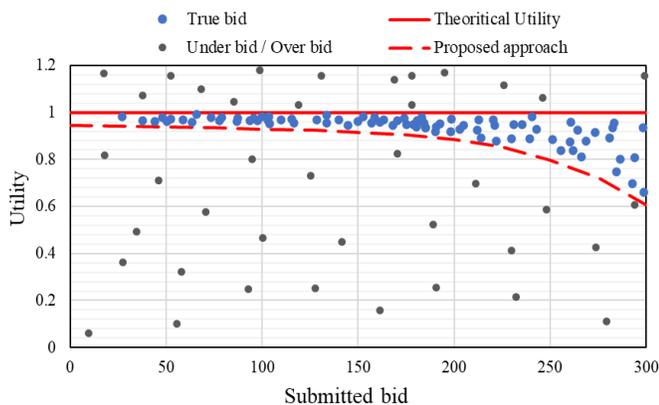

Fig. 6. Rationality analysis in forms of achieved utility under true bids and misreported bids

### E. Successful rate

We evaluate the successful rate of the three algorithms in Fig. 7. As the number of UEs' requests varies from 2,000 to 8,000, the successful rate declines. This is because the number of successfully assigned tasks remains almost constant under the capacity limitations of the ESs, despite the increase in total requests. Additionally, our approach algorithm achieves success rates that are 13 percent and 18 percent higher than the RACO and the RALA algorithms, respectively.

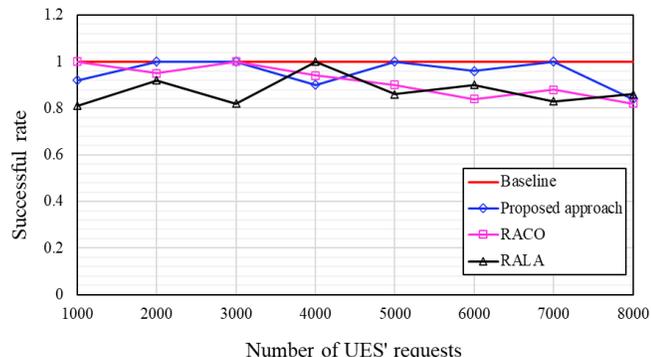

Fig. 7. Successful rate for various numbers of UEs' requests at the SHARCNET setting

### F. Computational efficiency

Fig. 8 depicts the computational efficiency of all methods under different numbers of offloaded tasks. As the number of UEs and tasks increases, the execution time of all approaches also increases. When there are 50 UEs and 5,000 tasks, the execution time of our method is less than 47 ms, which is significantly shorter than the RACO, with an execution time of 68 ms, and the RALA, with an execution time of 73 ms.

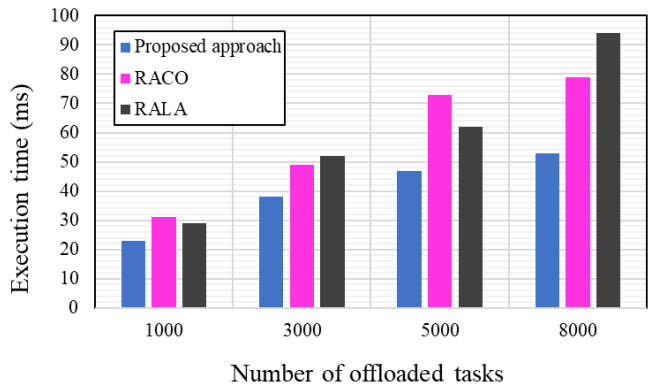

Fig. 8. Computational efficiency of all methods for different number of tasks under DAS-2 log file

### G. Convergence to the equilibrium

Fig. 9 analyzes the convergence to the auction equilibrium for all methods by monitoring the stability of cost values over successive auction cycles. As expected, all methods converge to the equilibrium. The results indicate that the proposed method converges to the equilibrium faster than the RACO and the RALA auction methods, which exhibit slower and less stable convergence patterns. Specifically, as the number of UEs reaches 180 and the UEs' requests for offloading climb to 5,000, our method consistently stabilizes in utility value within fewer iterations, see Fig.10. This convergence validates the mechanism's robust design, ensuring truthful bidding and optimal offloading decision.

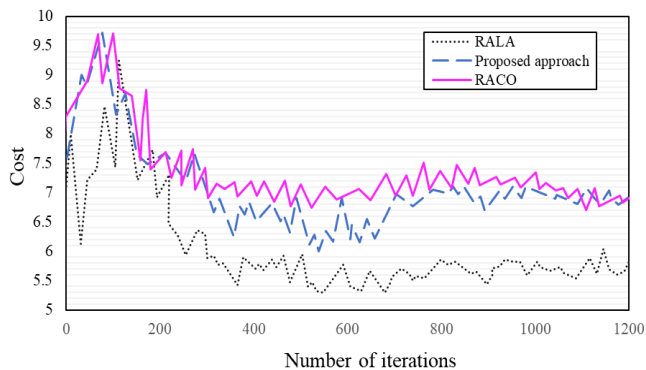

Fig. 9. Cost convergence for different numbers of iteration

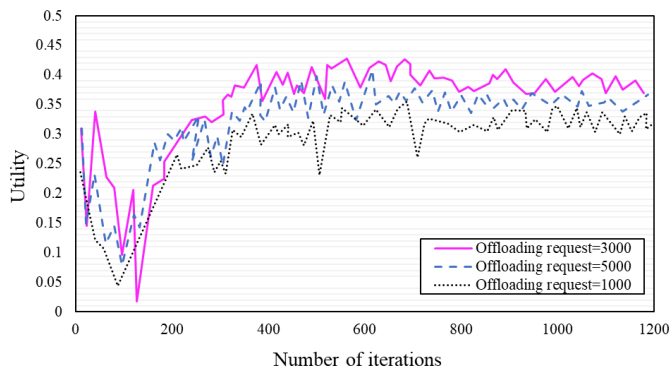

Fig. 10. Utility convergence of the proposed method under various numbers of offloading request

## VI. Conclusion

In this study, we proposed a task-offloading approach for delay-sensitive applications in edge-computing IoT environments. Our method takes into account a tradeoff between the delay constraint of tasks and offloading cost in task offloading decisions. We designed a demand and supply model to align UEs' computational demands with available ESs' resources as supply, ensuring supply meets user demands without overloading servers. To consider the preferences and requirements of both parties, UEs and ESs, and model the inherent competition behavior of the UEs over available computational resources, we formulated the multi-agent task offloading as a VCG auction-based game. By developing a sharing-incentive mechanism, we incentivized both parties to participate in the auction. The proposed game maximizes social welfare rather than individual payoff, where the collective utility of all users is prioritized, leading to a more equitable distribution of resources and improved performance for the network. Comprehensive simulations demonstrated the superiority of our method in truthfulness, envy-free resource allocation, and sharing incentive fulfillment. For future work, we plan to utilize machine learning techniques to predict user demand and resource availability, speed up the computation, and make the system scalable for larger networks.